%% file: main.tex
\documentclass{sigchi}

\CopyrightYear{2020}
\setcopyright{acmlicensed}

\toappear{
\scriptsize 
Permission to make digital or hard copies of all or part of this work 
for personal or classroom use is granted without fee provided that 
copies are not made or distributed for profit or commercial advantage 
and that copies bear this notice and the full citation on the first 
page. Copyrights for components of this work owned by others than ACM 
must be honored. Abstracting with credit is permitted. To copy 
otherwise, or republish, to post on servers or to redistribute to 
lists, requires prior specific permission and/or a fee. Request 
permissions from permissions@acm.org. \\
\emph{CHI '20, April 25--30, 2020, Honolulu, HI, USA.} \\
\copyright~2020 Association of Computing Machinery. \\
ACM ISBN 978-1-4503-6708-0/20/04\ ...\$15.00. \\
http://dx.doi.org/10.1145/3313831.3376715
}

\usepackage{balance}       
\usepackage{graphics}      
\usepackage[T1]{fontenc}   
\usepackage{txfonts}
\usepackage{mathptmx}
\usepackage[pdflang={en-US},pdftex]{hyperref}
\usepackage{color}
\usepackage{booktabs}
\usepackage{textcomp}
\usepackage{url}
\usepackage{dirtytalk}
\usepackage{nicefrac}
\usepackage{multirow}
\usepackage{multicol}
\usepackage{makecell}

\usepackage{microtype}        
\usepackage{ccicons}          

\usepackage{todonotes}
\usepackage{comment}
\usepackage{xspace}

\newcommand{\kenneth}[1]{{\small\color{blue}{\bf\xspace#1 -Kenneth}}}
\newcommand{\yang}[1]{{\small\color{orange}{\bf\xspace#1 -CY}}}

\newcommand{\system}{Heteroglossia\xspace}

\def\plaintitle{\system: In-Situ Story Ideation with the Crowd}

\def\emptyauthor{}
\def\plainkeywords{Authors' choice; of terms; separated; by
  semicolons; include commas, within terms only; this section is required.}

\makeatletter
\def\url@leostyle{%
  \@ifundefined{selectfont}{
    \def\UrlFont{\sf}
  }{
    \def\UrlFont{\small\bf\ttfamily}
  }}
\makeatother
\def\pprw{8.5in}
\def\pprh{11in}

\definecolor{linkColor}{RGB}{6,125,233}
\hypersetup{%
  pdftitle={\plaintitle},
  pdfauthor={\emptyauthor},
  pdfkeywords={\plainkeywords},
  pdfdisplaydoctitle=true, 
  bookmarksnumbered,
  pdfstartview={FitH},
  colorlinks,
  citecolor=black,
  filecolor=black,
  linkcolor=black,
  urlcolor=linkColor,
  breaklinks=true,
  hypertexnames=false
}

\begin{document}

\title{\plaintitle}

\numberofauthors{1}
\author{%
  \alignauthor{Chieh-Yang Huang, Shih-Hong Huang, and Ting-Hao (Kenneth) Huang\\
  \affaddr{College of Information Sciences and Technology}\\
    \affaddr{Pennsylvania State University, University Park, PA, USA}\\
    \email{\{chiehyang,~szh277,~txh710\}@psu.edu}
}
}



\maketitle

\begin{abstract}
\input{abstract.tex}

\end{abstract}

\ccsdesc[500]{Information systems~Crowdsourcing}
\ccsdesc[500]{Human-centered computing~Human computer interaction (HCI)}
\ccsdesc[300]{Human-centered computing~User studies}

\keywords{Crowdsourcing; Creative Writing; Ideation; Role Play; Story}

\printccsdesc

\section{Introduction}
\input{intro.tex}

\section{Related Work}
\input{related_work.tex}

\section{Heteroglossia System}
\input{system.tex}

\section{Study 1: The Effects of Role Play Strategy}
\input{experiment-role-play.tex}

\section{Study 2: Deployment with Creative Writers}
\input{deployment.tex}

\section{Discussion}
\input{discussion.tex}

\section{Conclusion and Future Work}
\input{conclusion-and-future-work.tex}

\section{Acknowledgements}
We thank Sooyeon Lee, Tiffany Knearem, Chi-Yang (Ethan) Hsu, Lisa Yu, and Frank Ritter for their valuable feedback and help.
We also thank the workers on Mechanical Turk who participated in our studies.

\balance{}

\bibliographystyle{SIGCHI-Reference-Format}
\bibliography{main}

\end{document}

%% file: abstract.tex

Ideation is essential for creative writing.
Many authors struggle to come up with ideas throughout the writing process, yet modern writing tools fail to provide on-the-spot assistance for writers when they get stuck.
This paper introduces \system, an add-on for Google Docs that allows writers to elicit story ideas from the online crowd using their text editors.
Writers can share snippets of their working drafts and ask the crowd to provide follow-up story ideas based on it.
\system employs a strategy called ``\textit{role play}'', where each worker is assigned a fictional character in a story and asked to brainstorm plot ideas from that character's perspective.
Our deployment with two experienced story writers shows that \system is easy to use and can generate interesting ideas.
\system allows us to gain insight into how future technologies can be developed to support ideation in creative writing.

%% file: intro.tex
Storytelling is one of the oldest known human activities~\cite{wiessner2014embers}.
People engage in storytelling to communicate, teach, entertain, establish identity, or simply relate to each other in meaningful ways~\cite{roemmele2018neural}.
Storytelling is important, but writing a good story is a challenging and complicated task, and many creative writers struggle to come up with ideas throughout the process.
Roland Barthes said: ``A creative writer is one for whom writing is a problem.'' 
Despite this common experience, research into technological writing support does not have much to say to help story writers.
Writing support systems have long been focused on business and technical writing.
Researchers have created systems that can automatically generate follow-up text in an auto-complete manner~\cite{Clark:2018:CWM:3172944.3172983}; decompose and recompose complicated writing tasks~\cite{kittur2011crowdforge}; outsource writing jobs to online crowds~\cite{bernstein2010soylent,hahn2016knowledge}, collaborators~\cite{teevan2016supporting}, or writers themselves~\cite{teevan2014selfsourcing}; 
or even allow the user to write a paper solely using a smartwatch~\cite{nebeling2016wearwrite}.
However, these prior works were largely developed and tested for producing technical reports~\cite{teevan2016supporting,teevan2014selfsourcing,nebeling2016wearwrite}, Wikipedia-like essays~\cite{kittur2011crowdforge}, or business documents~\cite{greer2016introduction}, rather than short stories or novels.
One of the few exceptions is the work done by Kim {\em et al.}, who created Ensemble~\cite{kim2014ensemble} and Mechanical Novel~\cite{kim2017mechanical}.
These two systems pushed the boundaries of collaborative story writing, but did not focus on helping creative writers, who mostly write alone~\cite{maxey_2007,geiger2019creativity,redvall2012european}.


\begin{figure}[t]
    \centering
    \includegraphics[width=1\columnwidth]{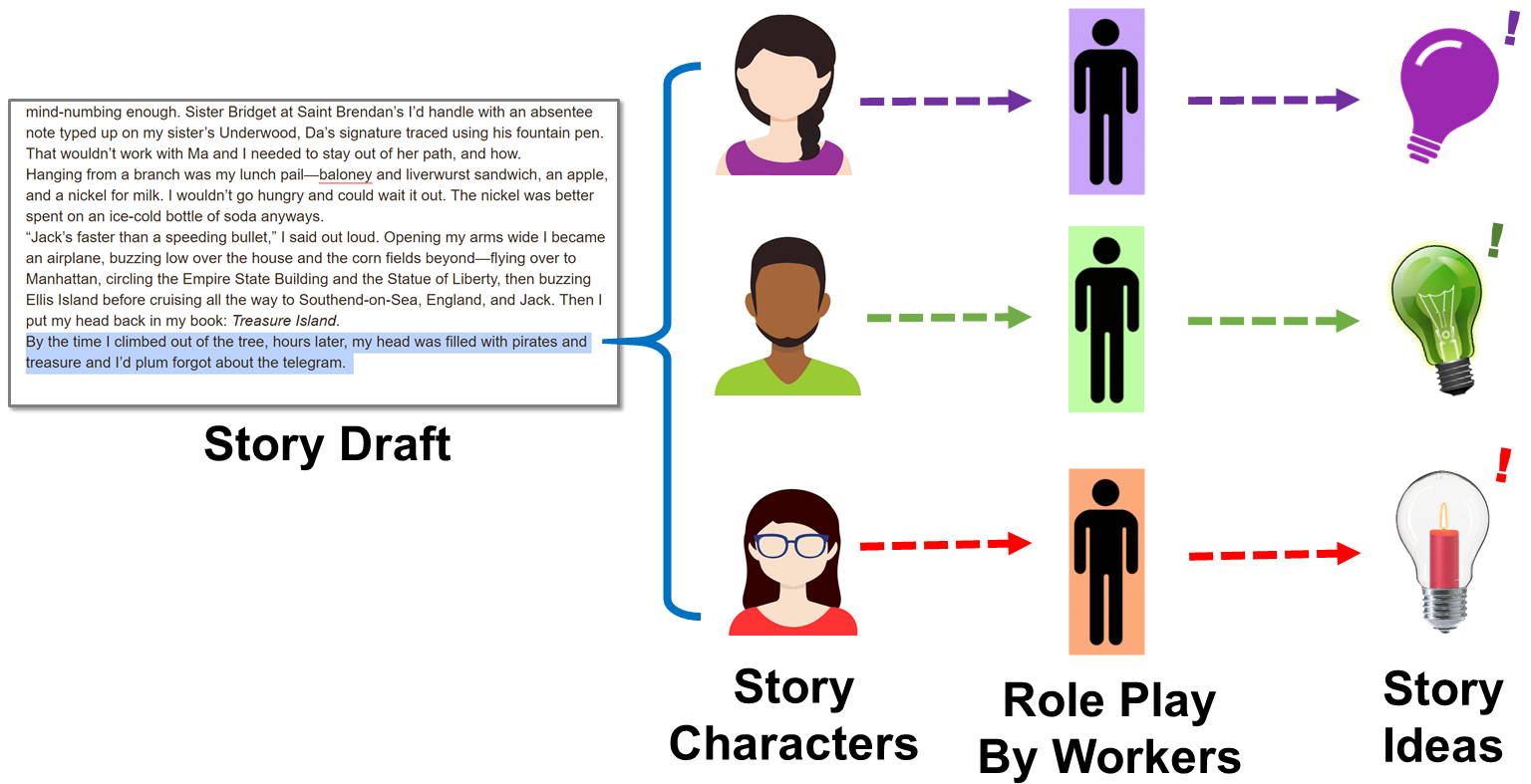}
    \caption{The overview of the role play ideation strategy. In order to obtain follow-up story ideas for a working draft, we recruit a group of crowd workers and ask them to imagine they were a character in the given story. Each worker is instructed to assume the role of a character in the story and generate plot ideas from this character's perspective.}
    \label{fig:overview}
    \vspace{-1pc}
\end{figure}

This paper introduces \textbf{\system}\footnote{Heteroglossia: a diversity of voices, styles of discourse, or points of view in a literary work and especially a novel~\cite{merriam-webster}.}, a crowd-powered system that allows writers to elicit story ideas simply using their text editors.
Figure~\ref{fig:overview} overviews the \system system, which we built as an add-on for Google Docs.
A writer can select a part of a story draft as the prompt and ask the online crowd to provide follow-up story ideas based on it.
\system employs an ideation strategy called ``\textbf{role play},'' where each worker is assigned a fictional character in the story and asked to brainstorm plot ideas from the character's perspective.
This work is motivated by the fact that role-playing and acting traditionally have had a role in the creative writing process~\cite{daiute1989play}.
Some professional novelists also use role play to help writing.

Also relevant is the well-known ``six hats'' method, which asks people to wear metaphorical hats representing different thinking perspectives~\cite{de2017six}.
Teevan {\em et al.} proposed to use the six hats schema to assign different thinking roles to the authors themselves in order to promote self-reflection from different angles~\cite{teevan2017bringing}.
Chou {\em et al.} showed that perspective-taking can release the fixation and thus help generate new ideas by asking people to imagine themselves in different roles involved in different activities~\cite{chou2017finding}. 
We developed our work based on these inspiring prior works. 
In this paper, we first use a set of controlled experiments to quantitatively illustrate the property of the role play ideation strategy, and then overview our system deployment with two experienced creative writers.
We believe \system allows us to gain insight into how future technologies can be developed to support creative writing.

%% file: related_work.tex
This work is related to 
{\em (i)} crowd ideation, 
{\em (ii)} crowd writing,
{\em (iii)} supporting creative writing, and
{\em (iv)} crowd feedback.

\subsection{Crowd Ideation}
Prior work has used the online crowd as a source of new ideas, primarily for problem-solving and product design.
Chan {\em et al.} introduced IdeaGens, an ideation system where a group of workers proposes ideas in real-time and the expert monitors the incoming ideas and provides instant feedback to the crowd~\cite{chan2016improving}.
Yu {\em et al.} explored using a schematic representation for the target design problem to guide the crowd to ``think outside of the box''~\cite{Yu:2016:EBT:2818048.2820025}.
Online crowds were also used to provide real-time creative input during early-stage design activities~\cite{andolina2017crowdboard}. 

\subsection{Crowd Writing}
\system also builds upon the work in crowd writing, which aims to allow a group of people, including experts and non-experts, to work together to write an article.
Many crowd writing projects focused on decompose and recompose complicated writing tasks.
For example, the Knowledge Accelerator used a complex workflow where each worker contributes small amounts of effort to synthesize online information, generating a Wikipedia-like article for open-ended questions~\cite{hahn2016knowledge}.
Soylent used a Find-Fix-Verify workflow to allow crowd workers to identify problems in a draft, propose solutions, and select the best solution for each identified problem~\cite{bernstein2010soylent}.
MicroWriter decomposed the task of writing into three subtasks: idea generation, labeling, and writing~\cite{teevan2016supporting}.
Meanwhile, some other work has pushed the boundaries of the classic workflow approach for crowd writing.
For example, WearWrite explored using wearable devices, such as smart watches, to guide a group of crowd workers to write articles~\cite{nebeling2016wearwrite}.
Agapie {\em et al.} explored using local crowds to generate event reports~\cite{agapie2015crowdsourcing}. 
However, these projects all focused on business or technical writing, rather than creative writing.

One of the few exceptions is the work done by Kim {\em et al.}, who created Ensemble~\cite{kim2014ensemble} and Mechanical Novel~\cite{kim2017mechanical}.
Ensemble is a volunteer-based collaborative story competition platform where Leaders set high-level creative goals and constraints for a story and Contributors participate in low-level tasks, such as drafting, commenting, and voting. Mechanical Novel embodies a more organic workflow, the ``Reflect-and-Revise loop,'' that allows crowd workers to revisit and revise their writing goal. These works pushed the boundaries of creative writing and helped to answer why collaborative novels at a scale similar to that of Wikipedia do not exist. 

While collaborative writing has opened new possibilities, most writers still write alone. 
Professional novelists write alone~\cite{maxey_2007}, freelance writers write alone~\cite{geiger2019creativity}, and, even within an industry with a collaborative culture, many TV screenwriters still write alone~\cite{redvall2012european}. 
Our goal is to assist creative writers, who often write alone, without drastically changing the way they work.

\subsection{Supporting Creative Writing}
A few researchers have developed technologies to support creative writing. 
Most of them focused on lower-level text generation or proofreading.
For example, the Creative Help system used a recurrent neural network model to generate suggestions for creative writing~\cite{roemmele2015creative}.
The Scheherazade system was developed for interactive narrative generation~\cite{li2015scheherazade}.
InkWell produced stylistic variations on texts to assist creative writers~\cite{gabriel2015inkwell}.
More recently, Clark {\em et al.} studied machine-in-the-loop story writing and suggested that machine intervention should balance between generating coherent and surprising suggestions~\cite{Clark:2018:CWM:3172944.3172983}. 

\subsection{Crowd Feedback Systems}
Researchers have also attempted to use online crowds to generate critiques and feedback.
Xu {\em et al.} created Voyant, a system that used non-expert crowd workers to generate structured feedback on visual designs~\cite{xu2014voyant}.
Their classroom study further demonstrated the effectiveness of using crowd feedback in the design process~\cite{xu2015classroom}.
As for visual designs, Luther {\em et al.} also created CrowdCrit, a system that aggregated multiple critiques from non-expert crowd workers~\cite{luther2015structuring}.
Luther showed in experiments that the critiques generated by CrowdCrit could help designers improve their design processes.
On the other hand, some other researchers focused on generating writing feedback. For example, Huang {\em et al.} used workers from Amazon Mechanical Turk (Mturk), who are often fluent in English, to provide structural feedback for ESL writing~\cite{huang2017supporting}.

%% file: system.tex
\begin{figure*}[t]
    \centering
    \includegraphics[width=.87\textwidth]{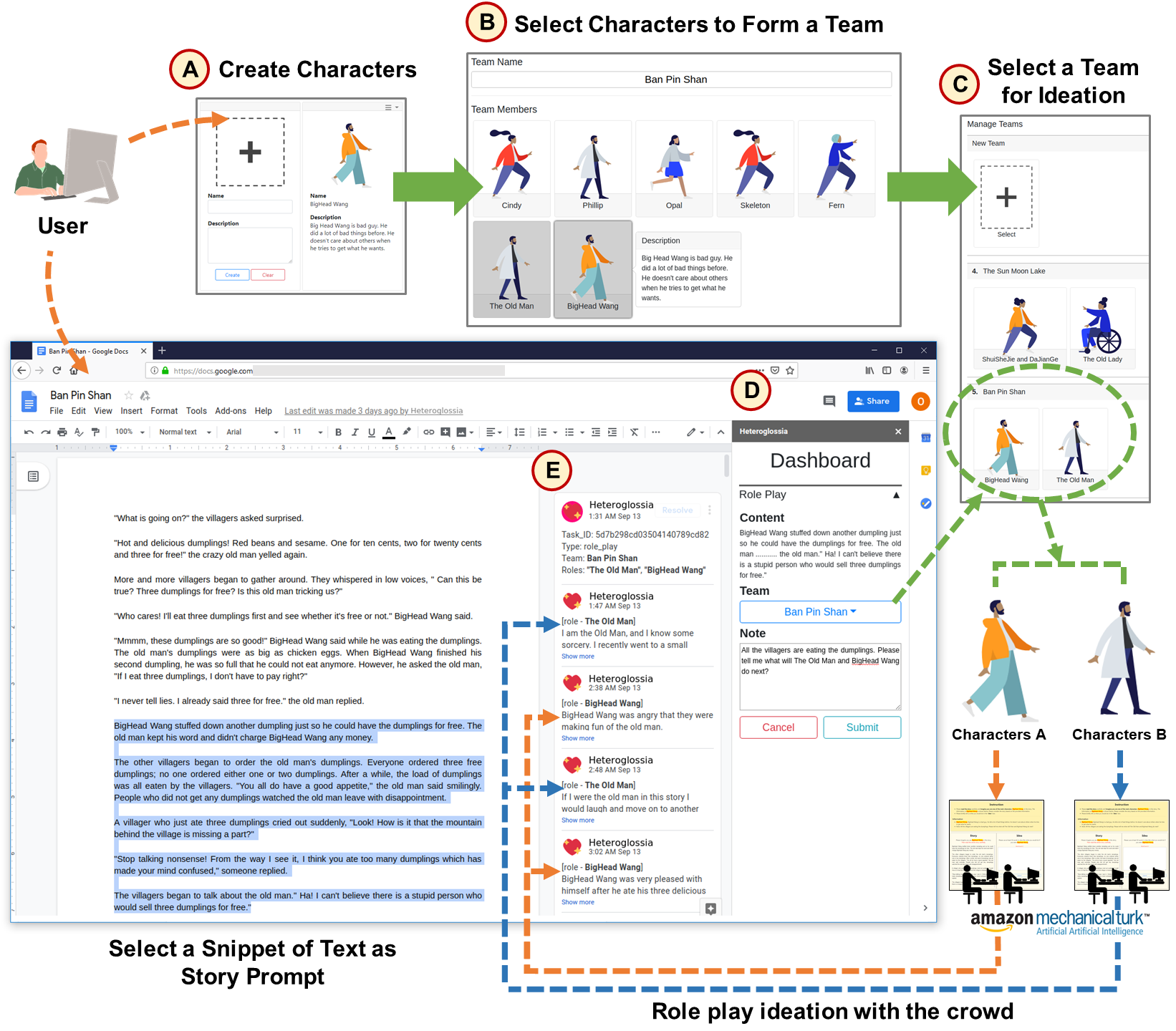}
    \caption{The system overview of \system. \system incorporates a web site to manage information and a Google Docs add-on for writing. Users start by creating characters (A) and forming teams of characters (B and C) on the \system website. After setting up the characters and teams, users start writing their own story in Google Docs. When they get stuck, users can select a story snippet to initiate an ideation task through \system (D) and acquire follow-up story ideas (E).}
    \label{fig:ui}
    \vspace{-0.4cm}
\end{figure*}


\system incorporates a web site to manage information and a Google Docs add-on for writing.
Figure~\ref{fig:ui} shows the screenshot of each page of \system website.
Users start by creating characters (Figure~\ref{fig:ui}A) and forming teams of characters (Figure~\ref{fig:ui}B and \ref{fig:ui}C) on the \system website.
After setting up the characters and teams, users start writing the story in Google Docs.
When they get stuck, users can select a story snippet to initiate an ideation task through \system (Figure~\ref{fig:ui}D) and acquire follow-up story ideas (Figure~\ref{fig:ui}E).

\subsection{Creating Characters}
Figure~\ref{fig:ui}A shows that to create a new character, users specify an image, name, and description.
Notice that only the name and description will be shown to workers.
Users can provide a detailed setting for a character in the description, such as inner goal, outer goal, and personality, to help workers understand the story background and come up with new story ideas.
Editing and deleting an existing character can be done through the setting button in the upper right corner.

\subsection{Forming a Team of Characters}
A team represents a group of characters used in role play ideation.
Figure~\ref{fig:ui}B shows the interface of editing a team.
Available characters are listed in the ``Team Members'' block with selected characters highlighted in gray.
Existing teams are listed row by row, as shown in Figure~\ref{fig:ui}C, and are available to edit and delete.

\subsection{In-Situ Story Ideation with the Crowd}
\system adopts Google Docs as its main platform, taking advantage of its convenient and well-maintained \textit{comment} functionality. 
The Google Docs add-on is implemented in Google Apps Script.
As shown in Figure~\ref{fig:ui}D, users can easily initiate a new ideation task in \system by 
{\em (i)} selecting a snippet of text on the Google Docs, which \system uses as the story prompt (the ``Content'' field), 
{\em (ii)} picking up a suitable team for role-play (using the ``Team'' button), and 
{\em (iii)} writing down guiding information to workers in the ``Note'' field (optional).

After submitting a new ideation task, \system will automatically generate corresponding pages and create Human Intelligence Tasks (HITs) on MTurk to recruit workers.
At the same time, \system will create a new comment that overviews this ideation task ({\em e.g.,} which team is used, which characters are in this team) and associate it with the selected text in Google Docs.
The first comment in Figure~\ref{fig:ui}E is an example. 
Upon receiving a worker's assignment, \system will present the story idea to the user as a reply to the initial overview comment. Figure~\ref{fig:ui}E shows four different story ideas from both characters. 
Since Google doesn't provide enough support to manipulate comments, \system's comment function is implemented by Google Doc API, Google Drive API, and a bot built with Selenium.

\subsection{Worker Interface}
\begin{figure}[t]
    \centering
     \includegraphics[width=.85\columnwidth]{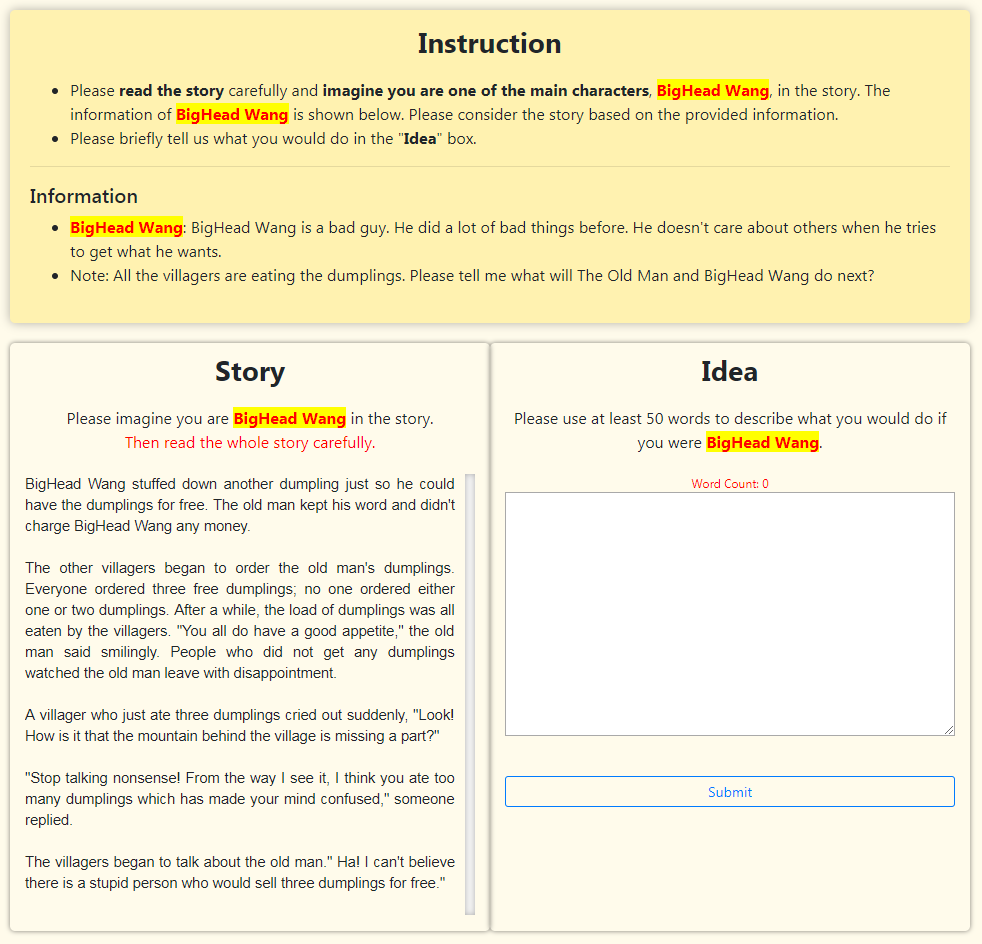}
    \caption{\system worker interface. The interface contains an instruction pane, a story pane, and an idea pane.}
    \label{fig:worker_ui}
    \vspace{-0.35cm}
\end{figure}

The worker interface contains an instruction pane, a story pane, and an idea pane, as shown in Figure~\ref{fig:worker_ui}.
User-defined information, a character description, and a task note are given in the instruction pane.
Workers are required to read the story prompt in the story pane and enter an idea in the idea pane.
To emphasize the role-play strategy, we display the character name in all three panes highlighted in red.
To prevent workers from behaving differently from our expectations, three rules are implemented on the worker interface: a 30-second time lock for HIT submission, a reach-to-the-bottom check for the story prompt, and a prohibition of copy-paste functionality in the idea pane.

\subsection{Dynamic Payment for Workers}
We proposed a formula to dynamically estimate working time and set up corresponding payment for workers.
The estimation is based on two factors: reading comprehension and writing.
The average reading speed of English native speakers is 200-300 words per minute with reasonable comprehension when using LCD monitors~\cite{siegenthaler2012reading}. 
We empirically estimated that writing a fifty-word-long story idea in \system takes approximately 5-6 minutes.
Aiming at providing a \$10 hourly wage, we implemented the formula as follows:
\begin{align}
\begin{split}
    Cost(HIT) &= Cost(Reading) + Cost(Writing)  \\
              &= \$(\nicefrac{\# words}{1000}) + \$1.0
\end{split}
\end{align}
where $\# words$ refers to the word count of the story prompt.
\system then creates HITs with the reward dynamically computed according to the designed formula.

%% file: experiment-role-play.tex
The goal of \system is to provide inspiring ideas to creative writers, especially when they get stuck during writing.
\system particularly uses an ideation strategy called ``role play.''
To understand the effects of the role-play strategy and inform the design of \system, we conducted two sets of experiments.
We would like to answer these two questions that are motivated by literature: 
{\em (i)} can the role play strategy produce more useful story ideas? and 
{\em (ii)} what are some trade-offs of using this strategy?

\subsection{Role Play Produces Semantically-Far Story Ideas}
\input{semantic-distance.tex}

\subsection{Trade-offs Between Task Structures and Creativity}

\input{trade-offs.tex}

%% file: semantic-distance.tex
Per Chan {\em et al.}~\cite{chan2017semantically,chan2018best}, when a creator reaches an impasse, ideas that are semantically far from current working ideas are more helpful than those that are nearer.
Chan's work is powered by the Search for Ideas in Associative Memory (SIAM) theory~\cite{nijstad2006group} and verified with crowd ideation experiments.
SIAM~\cite{nijstad2006group} assumes that idea generation is proceed in two stages, knowledge activation and idea production. 
In the first stage, an image will be retrieved according to the problem. 
The given image is assumed to have several features that is then used to generate ideas. 
Chan {\em et al.}~\cite{chan2017semantically,chan2018best} showed that in the idea production stage, relevant stimulations help generate more ideas. 
However, after exhausting the related ideas, semantically far stimulations would help people to change the category of images and thus generate more ideas. 
Applying Chan's conclusion to our system says that the theoretical prerequisites for resolving writer's block are to come up with story ideas that with greater semantic distance from the current working draft.
In this subsection, we conducted a set of experiments to examine if role-play strategy results in semantically distant ideas.

\begin{figure}
    \centering
    \includegraphics[width=1.0\linewidth]{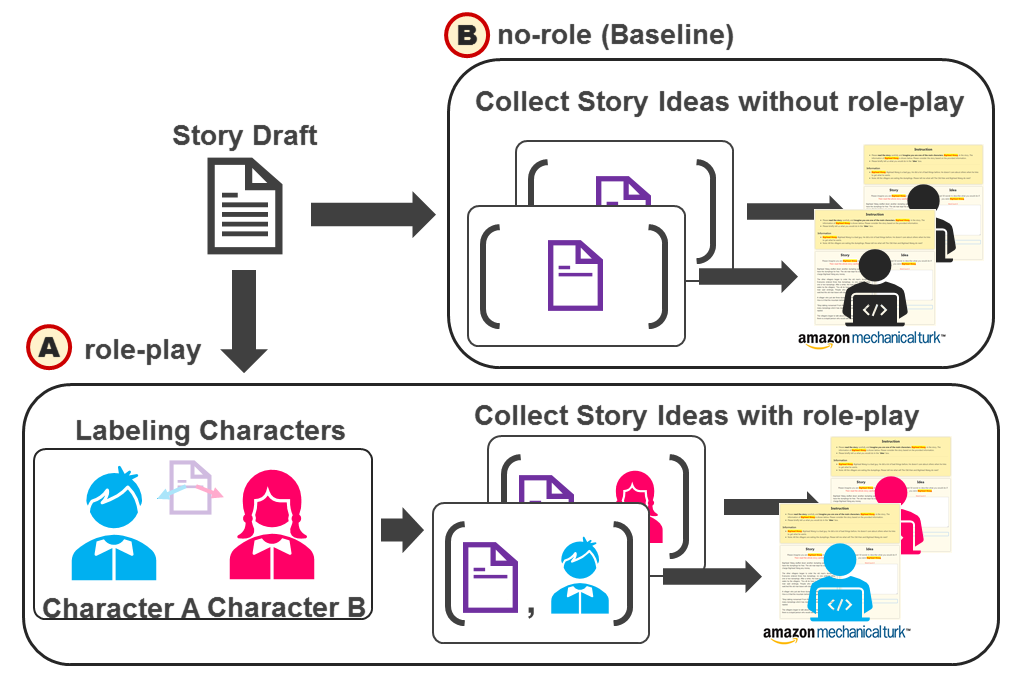}
    \caption{The overview of the Study 1. Two settings of ideas were collected: (A) role-play and (B) no-role. Notice that the same number of ideas was collected for a fair comparison.}
    \label{fig:pilot_study}
    \vspace{-0.3cm}
\end{figure}


\textbf{Pilot Study:}
The overview procedure of the pilot study is shown in Figure~\ref{fig:pilot_study}.
We first conducted a pilot study using five Taiwanese folk stories\footnote{Taiwanese folk stories were retrieved from TaiwanDC (https://www.taiwandc.org/folk.htm). We used ``The Legend of Sun-Moon Lake,'' ``The Legend of Muddy River,'' ``The Lake of the Sisters and the Tree Brothers,'' ``Ban Pin Shan,'' and ``The Tigress Witch (Hoko Po)''}.
These stories are unfamiliar to many crowd workers in order to simulate the workers' sense of freshness when reading the stories. 
We then took the first 30\% and first 60\% of each story (based on word count) to simulate a writer's working story drafts.
For each of these ten ($5\times2$) story drafts, we manually labeled the fictional characters in the story.
Each story contained two or three characters.
For each \textit{(story draft, character)} tuple, we recruited five workers from MTurk to read the story draft and provide a follow-up story idea in free text from the character's perspective.
For comparison, we also recruited ($\#character\times5$) workers for each draft without assigning any characters, and asked workers to write story ideas.
We paid \$0.5 for $30\%$ story drafts and \$0.8 for $60\%$ story drafts.
Each worker was allowed to work on each story once, {\em i.e.}, five was the maximum.
In total, 105 workers participated in our pilot study.

The pilot study leads to four main findings:

\begin{enumerate}
    \item \textbf{Paragraph-level semantic distance measurement is needed.} Chan {\em et. al}'s work either focused on short text~\cite{chan2017semantically} or large-scale ideation data~\cite{chan2018best}. However, in \system, the prompts and ideas will be in paragraphs, necessitating a paragraph-level semantic distance measurement. We introduce using doc2vec to measure semantic distances automatically, which we will describe in later subsections.
    \item \textbf{The role play strategy resulted in longer semantic distance.} Using doc2vec (Wiki), we estimated the semantic distance between the story draft and ideas. The ideas that came from workers with a role measured 0.490 and without a role measured 0.468.
    \item \textbf{There's a need to know where the writer actually got stuck.} The story drafts used in the pilot study were not actually interrupted where the writer got stuck. Some of the drafts were even segmented at where the follow-up plot is straightforward.
    \item \textbf{Story prompts could miss critical context.} Workers sometimes provided ideas that conflicted with the core character setting because they simply did not know it. Prior works have demonstrated that maintaining context is critical for designing efficient crowdsourcing workflows. We will allow users to supplement the background information of each character in \system. 
\end{enumerate}


\begin{table*}[t]
    \centering
    \small
    \begin{tabular}{lrrrrrrrrr}
    \toprule
         \textbf{Metric} & \textbf{GloVe} & \textbf{D2V-Wiki} & \textbf{D2V-News} & \textbf{D2V-Story} & \textbf{ST} & \textbf{S-ST Mean} & \textbf{S-ST Min} & \textbf{S-ST Median} \\ \midrule
        \textbf{$\rho$} & \textbf{-0.132} & 0.159 & 0.015 & \textbf{-0.153} & -0.021 & 0.202 & 0.129 & 0.178 \\ \midrule
        \textbf{$\tau$} & \textbf{-0.053} & 0.075 & -0.011 & \textbf{-0.128} & 0.080 & 0.192 & 0.080 & 0.171 \\
    \bottomrule
    \end{tabular}
    \caption{Correlation between human-rated relevance and automatic evaluation scores. Note that a good semantic distance indicator should negatively correlate with relevance scores. Only GloVe and Doc2Vec-Story generate scores that are negatively correlated to relevance in both Pearson ($\rho$) and Kendall ($\tau$) correlation coefficients. Doc2Vec-Story is a stronger indicator than GloVe because it yields higher correlation scores.}
    \label{tab:correlation}
    \vspace{-0.15cm}
\end{table*}

\textbf{Data Preparation:}
In response to the need of knowing where the writer actually got stuck, we acquired the data collected by the Creative Help system~\cite{roemmele2015creative} for further study.
Creative Help is an online writing application where users can freely write stories.
When the writer explicitly requests for help, the system automatically generates suggestions for the next sentence in a story. 
Users can modify, delete, or adopt the suggestions.
We considered a user request in Creative Help as a strong signal indicating the writer gets stuck.
Creative Help collected 1,078 stories during its deployment between 2015 to 2018.
A story contains one or more ``instances'', each represents a help request sent from the writer to the Creative Help system.
We removed stories with fewer than 20 sentences or fewer than three requests, resulting in a total of 107 stories.
Each story on average contains 37.8 (SD=17.4) sentences and 510.9 (SD=252.8) words.
We further removed stories that are obviously copied from the Internet or generated entirely by Creative Help without any human-written parts.
One co-author and one collaborator then labeled the characters that appeared in each story, respectively.
Only 14 stories, whose characters were totally agreed by two annotators, were used in the following experiment.
We segmented each story at the \textit{second last} request and used it as the story prompt.

\textbf{Story Ideation Using Role Play Strategy:}
We used the same interface as \system (Figure~\ref{fig:worker_ui}) to collect story ideas.
Identical to the pilot study, two conditions, [role] and [no-role], were subject to experiment.
In the [role] condition, workers were instructed to imagine that they were one of the main characters in the story and provide a follow-up plot idea in free text.
In the [no-role] condition, we asked workers to provide ideas without any constraints.
A total of 330 stories were collected contributed by 101 workers.

\textbf{Human Evaluation:}
For each received story idea, we recruited another five workers from MTurk to rate the semantic distance to the story prompt.
We collected the rating scores using a 5-point Likert scale of agreement with the statement, ``This story idea is a \textit{relevant} follow-up of the original story prompt.'' (1 = Strongly Disagree, 5 = Strongly Agree.)
Table~\ref{tab:conceptual_distance} shows the results.
The story ideas collected in the [role] condition had an average relevance score of 3.869, while the ideas in the [no-role] condition had an average relevance score of 3.998. The difference is statistically significant (paired t-test, $p<0.05$, $N=14$.)
Namely, based on human evaluation, \textbf{the role play strategy generated semantically further ideas.}

Researchers used MTurk to evaluate creative works and resulted in high inter-annotator agreements~\cite{chan2017semantically,luther2015structuring}. 
However, some prior works also raised concerns about using non-experts to assess creative works such as graphic designs~\cite{jeffries2017cat} and poems~\cite{kaufman2008comparison}.
To further examine our findings, in the following subsection, we measured the semantic distance between text snippets using vector representations.

\textbf{Automatic Evaluation:}
Most of the automated distance measures require first representing text snippets as numeric vectors, called ``document representation.''
Previous studies~\cite{chan2017semantically}, where ideas were short pieces of text, simply summed up the corresponding GloVe vectors~\cite{pennington2014glove} and used cosine similarity for distance measurement.
However, in our study, the collected ideas on average contains 78.0 words (SD=27.8) and thus require a paragraph-based representation.
To this end, we experimented using the following six document representations to measure semantic distance:

\begin{enumerate}
    \item \textbf{GloVe:} We used pretrained GloVe vectors (glove.6B.300d)~\cite{pennington2014glove}. The document vector was obtained by summing up the corresponding word vectors and $1-cosine\ similarity $ was applied for distance measurement.
    \item \textbf{Doc2Vec (Wikipedia):} We used the Doc2Vec model~\cite{le2014distributed,lau2016empirical}, which was trained on the Wikipedia dataset (github.com/jhlau/doc2vec), and applied $1-cosine\ similarity$ as a function for distance measurement. 
    \item \textbf{Doc2Vec (News):} Same as \#2 but was trained on the Associated Press News dataset.
    \item \textbf{Doc2Vec (Story):} Same as \#2, but was trained on ROCStories~\cite{mostafazadeh2016corpus}.
    \item \textbf{Skip-thought Vector:} We used the pretrained skip-thought model~\cite{kiros2015skip} to encode the document and applied $1-cosine\ similarity$ as a function for distance measurement.
    \item \textbf{Sentence-level Skip-thought Vector (Mean):} We segmented a document into sentences first and encoded each sentence using the pretrained skip-thought model. Thus, the document can be represented as a set of vectors \textbf{V}$_{sst}$ = \{$v_1$, $v_2$, ..., $v_n$\}. When computing the distance between two sentence-level skip-thought vectors, we computed $1-cosine\ similarity$ among pairs. The \textbf{mean} over the distances of each pair was used as the distance measure.
    \item \textbf{Sentence-level Skip-thought Vector (Min):} Same as \#6, but used the \textbf{min} over the distances of every pair as the distance measure.
    \item \textbf{Sentence-level Skip-thought Vector (Median):} Same as \#6, but used the \textbf{median} over the distances of every pair as the distance measure.
\end{enumerate}

In order to evaluate how well these methods reflect human judgements, we calculated the correlation coefficients between the automatic score and human scores collected above.
Note that a good semantic distance indicator should \textit{negatively} correlate with relevance scores.
Table~\ref{tab:correlation} shows that only \textbf{GloVe and Doc2Vec-Story generate scores that are negatively correlated to human judgements of relevance} in both Pearson ($\rho$) and Kendall ($\tau$) correlation coefficients, where Doc2Vec-Story is a stronger indicator than GloVe because it yields higher correlation scores.

Finally, we used GloVe and Doc2Vec-Story to measure the semantic distance automatically.
Table~\ref{tab:conceptual_distance_auto} shows that both methods suggest the story ideas collected in the [role] condition had an longer semantic distance to the story prompt.
These automatic measurement can be used in \system to automatically rank the usefulness of received story ideas, or to filter out ideas that are abnormally similar to each other.


\begin{table}[t]
    \centering
    \small
    \addtolength{\tabcolsep}{-0.05cm}
    \begin{tabular}{lrrrrr}
    \toprule
    \multirow{2}{*}{} & \multicolumn{2}{c}{\textbf{No-Role}} & \multicolumn{2}{c}{\textbf{Role}} & \multicolumn{1}{c}{\multirow{2}{*}{\textbf{$d$}}} \\ \cmidrule{2-5}
                                      & \multicolumn{1}{c}{\textbf{Mean}} & \multicolumn{1}{c}{\textbf{95\% CI}} & \multicolumn{1}{c}{\textbf{Mean}} & \multicolumn{1}{c}{\textbf{95\% CI}} & \\ \midrule
    \textbf{Relevance} & *3.998 & [3.925, 4.070] & 3.869 &  [3.789, 3.948] & 0.89 \\
    \bottomrule
    \end{tabular}
    \addtolength{\tabcolsep}{0.05cm}
    \caption{Relevance of ideas, rated by human judges. The role play strategy (Role) generated semantically further ({\em i.e.,} less relevant) ideas. (*: $p<0.05$. Paired t-test. $N=14$. Cohen's $d$ reported as [no-role] - [role]. Large effect size: $\left | d \right |>0.8$.)}
    \vspace{-0.2cm}
    \label{tab:conceptual_distance}
\end{table}

\begin{table}[t]
    \centering
    \small
    \addtolength{\tabcolsep}{-0.05cm}
    \begin{tabular}{lrrrrr}
    \toprule
    \multirow{2}{*}{} & \multicolumn{2}{c}{\textbf{No-Role}} & \multicolumn{2}{c}{\textbf{Role}} & \multicolumn{1}{c}{\multirow{2}{*}{\textbf{$d$}}} \\ \cmidrule{2-5}
                                      & \multicolumn{1}{c}{\textbf{Mean}} & \multicolumn{1}{c}{\textbf{95\% CI}} & \multicolumn{1}{c}{\textbf{Mean}} & \multicolumn{1}{c}{\textbf{95\% CI}} & \\ \midrule
    \textbf{D2V-Story} & 0.840 & [0.819, 0.860] & 0.848 & [0.829, 0.866]  & -0.23 \\
    \textbf{GloVe} & 0.039 & [0.034, 0.044] & 0.045 & [0.037, 0.053]  & -0.47 \\
    \bottomrule
    \end{tabular}
    \addtolength{\tabcolsep}{0.05cm}
    \caption{Automatic evaluation metrics of semantic distance. Both D2V-Story and GloVe methods suggested that the story ideas collected in the [Role] condition had an longer semantic distance to the story prompt. (Cohen's $d$ reported as [no-role] - [role]. Small effect size: $\left | d \right |>0.2$.)}
    \label{tab:conceptual_distance_auto}
\end{table}

%% file: trade-offs.tex
\begin{table}[t]
\center
\small
\addtolength{\tabcolsep}{-0.07cm}
\begin{tabular}{lrrrrr}
\toprule
    \multirow{2}{*}{\textbf{Aspects}} & \multicolumn{2}{c}{\textbf{No-Role}} & \multicolumn{2}{c}{\textbf{Role}} & \multicolumn{1}{c}{\multirow{2}{*}{\textbf{$d$}}} \\ \cmidrule{2-5}
                                      & \multicolumn{1}{c}{\textbf{Mean}} & \multicolumn{1}{c}{\textbf{95\% CI}} & \multicolumn{1}{c}{\textbf{Mean}} & \multicolumn{1}{c}{\textbf{95\% CI}} & \\ \midrule
    \textbf{Legitimate} & **3.97 & [3.882, 4.052] & 3.81 & [3.735, 3.885] & 1.02 \\
    \textbf{Creative} & 3.66 & [3.534, 3.792] & 3.54 & [3.429, 3.654] & 0.53 \\
    \textbf{Interesting} & 3.63 & [3.493, 3.762] & 3.52 & [3.428, 3.608] & 0.50 \\
    \textbf{Willing-to-Read} & *3.60 & [3.473, 3.733] & 3.49 & [3.396, 3.587] & 0.51 \\
    \textbf{Surprising}  & *3.37 & [3.242, 3.503] & 3.23 & [3.105, 3.348] & 0.61 \\ \bottomrule
\end{tabular}
\addtolength{\tabcolsep}{0.07cm}
\caption{Trade-offs between task structures and creativity. Five human judges on MTurk rate each story ideas on the following five aspects, using a 5-point Likert scale: Legitimate, Creative, Interesting, Willing-to-Read, and Surprising. (*: $p<0.05$; **: $p<0.01$. Paired t-test. $N=14$. Cohen's $d$ reported as [no-role] - [role]. Medium effect size: $\left | d \right | > 0.5$. Large effect size: $\left | d \right | > 0.8$.)}
\label{tab:trade-off-v}
\vspace{-0.2cm}
\end{table}



Per Kim {\em et al.}~\cite{kim2014ensemble}, in collaborative story writing, task structures and creativity have some trade-offs.
Too little structure leads to ``unfocused, sprawling narratives'', and too much structure ``stifles creativity.''
The role play strategy enforces a schema of characters for ideation and could possibly sacrifice the quality or creativity of story ideas submitted by workers.
To understand the effect of role play strategy thoroughly, we conducted experiments to examine this possible trade-off.

For each story idea received in Study 1, we recruited five workers from MTurk to rate the quality of the idea in various aspects\footnote{Legitimate (``This story idea makes sense given the story prompt.''),
Creative (``This story idea is creative.''), 
Interesting (``This story idea is interesting.''),
Willing-to-Read (``I'm willing to read the final story that is written based on this story idea.''), and
Surprising (``This is a surprising story idea.'')}, using a 5-point Likert scale of agreement (1 = Strongly Disagree, 5 = Strongly Agree.) 
Table~\ref{tab:trade-off-v} shows the results, which echo Kim {\em et al.}'s observation that enforcing task structure could stifle creativity.
As for design implication, this trade-off should be made explicit to users and allow them to freely decide which strategy to use.

%% file: deployment.tex
\begin{figure*}[ht]
    \centering
    \includegraphics[width=.8\linewidth]{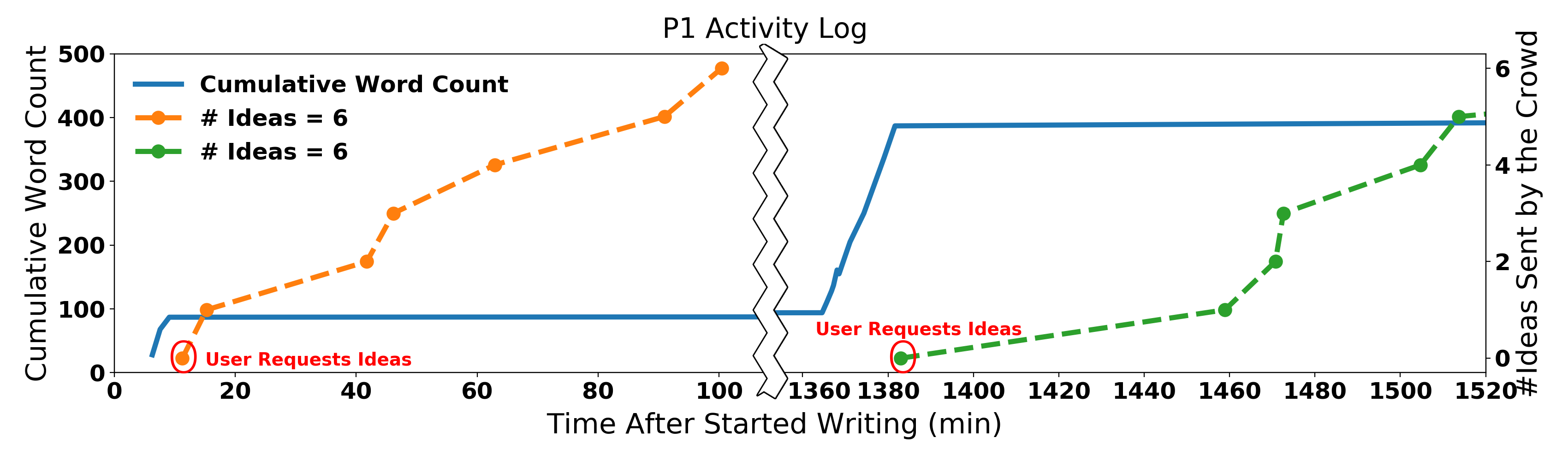}
    \caption{User activity logs of P1 shown in the cumulative word count with respect to time. When different numbers of characters were used, the resulting number of ideas varied. Therefore, the total number of ideas is shown in the legends. As we can see, participants usually requested ideas and paused writing. After hours, when most of ideas had appeared, they came back and resumed writing.}
    \label{fig:activity_log_1}
    \vspace{-1pc}
\end{figure*}

\begin{figure}[ht]
    \centering
    \includegraphics[width=1.0\linewidth]{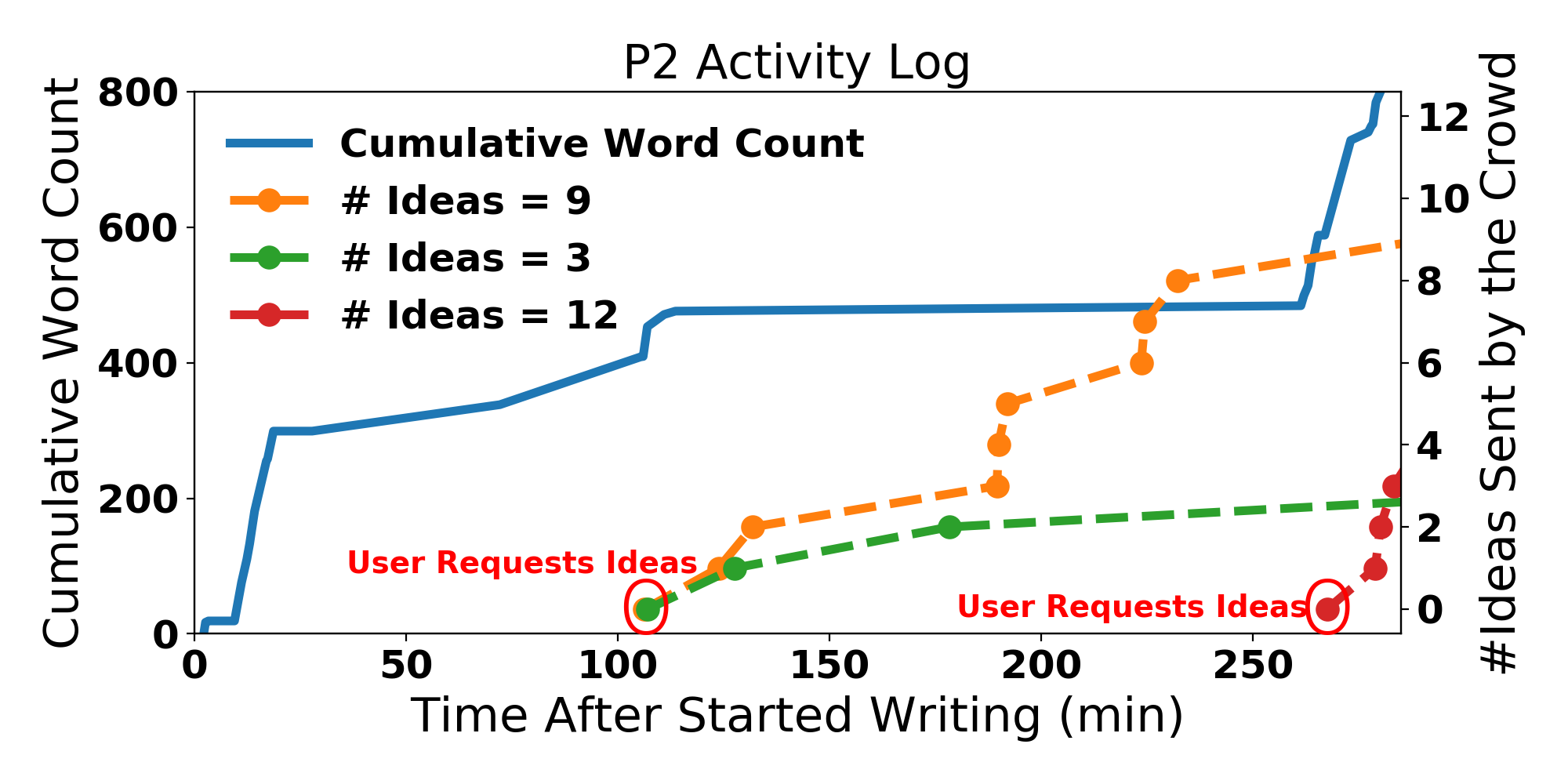}
    \caption{User activity logs of P2 shown in the cumulative word count with respect to time. P2 requested multiple tasks at the same time.}
    \label{fig:activity_log_2}
\end{figure}

To understand how writers would use \system, we conducted a three-day deployment study with two experienced creative writers and held pre-study and post-study interviews. 

\subsection{Participants}

Two experienced creative writers, P1 and P2, participated in our deployment study. Both participants are women and native speakers of American English.
They were recruited through our personal networks.
\textbf{P1} has been writing since she could pick up a pencil and has always written stories. 
She wrote a lot of fan fiction in middle school and high school.
P1 received an English minor in undergraduate school.
She started writing her own book in 2018 and is currently at the stage of pitching agents for publication.
P1 has also translated video games from Japanese into English.
P1 has never done technical writing, and her main genre focus is fantasy.
\textbf{P2} (age=32) has a minor in creative writing. 
She has participated in National Novel Writing Month (NaNoWriMo, www.nanowrimo.org) nine times and succeeded five times. (NaNoWriMo is an annual event where writers set goals to finish novels within the month.) 
P2 does not have publications but has written a great deal.
She has done technical writing before and mainly focuses on science fiction and science fantasy.
P2 uses Google Docs as her primary text editor.

\subsection{Study Protocol}
\label{sec:study-protocol}

Before the study, we had a semi-structured pre-study interview via Skype with the participants to understand their backgrounds, needs, and creative writing processes. 
At the end of the pre-study interview, one of the authors gave a brief tutorial of \system and demonstrated how to use the system.
Note that we explicitly informed the participants that \system is a crowd-powered system and that their stories would be viewed by online crowd workers.
In the study, the participants were asked to use \system to write a story of approximately 1,000 words.
We asked the participants to finish the story in a time span of three days, during which they needed to use \system's ideation function at least three times when writing their stories.
After the study, we had a semi-structured post-study interview via Skype with the participants to understand their experience and feedback.
The pre- and post-study interviews were both around thirty minutes long.
The audio was recorded and transcribed by the authors.
Each participant was compensated with \$50 after the post-study interview.
Table~\ref{tab:example} shows one example of crowd ideation created by P2.

\begin{table*}[t]
    \centering
    \small
    \begin{tabular}{p{.2cm}p{15.2cm}}
    \toprule
        \multicolumn{2}{p{15.5cm}}{\textbf{Story Prompt}} \\
        \multicolumn{2}{p{15.5cm}}{Detective Opal considered her seargant. Like all werewolves, Seargant Subwoofer looked like a normal person most of the time--he only went furry during the full moon--and so he could be contained in a Full Metal concrete barrier, and no harm would happen. Because of this, Werewolves were considered a class-A Supernatural. Other Supernaturals--who were a bigger threat to humanity--had classes B, C, D, and F. Even though Seargant Subwoofer was one of the least dangerous, I could see how people looked at him--it grated on him, I could tell.}  \\ \midrule
        \multicolumn{2}{p{15.5cm}}{\textbf{Detective Opal} - \textit{Detective Opal has a murder to solve in a fantasy world: and it's not as obvious who the killer is that she thought}} \\
            1. & I would have to clear the Sergeant's name even though I had a my doubts about his guilt. Finding his alibi should be easy enough as I clearly remember seeing a camera that eyed the only way in and out of his concrete self imposed prison. I can't deny that I how odd it was though when he asked to see the warrant for the tape I requested. \\
            2. & It bothered him so much that one day he hired a witch to remove the curse that was placed on him. The curse that turned him into a werewolf. After the curse was removed he aged significantly. Being a werewolf kept him young and now he was his real age. He didn't care though, because he was finally free from the curse. \\
            3. & Detective Opal is one of the unique supernaturals. She doesn't change every full moon, but rather has the powers to change between human and werewolf whenever she pleases. She generally uses the powers during investigation, smelling for scents, and chasing down perpetrators. Her supernatural abilities come in handy as a detective. \\ \midrule
        \multicolumn{2}{p{15.5cm}}{\textbf{Siren Eris} - \textit{Eris is a siren. She is being accused of murder, but she is innocent.}} \\
            1. & I was aware of how they could view me as a potential suspect. We were known to have our fights and screaming matches, but it never went any further. I was determined to set the record straight. I knew that the best way to do that was to provide them with my alibi. \\
            2. & I will approach them telling them the true story of how the crime happened. if there is a negative response, i will now tell them i wish to see sergeant Subwoofer and explain myself to him and beg him to show empathy on me. if it does not bring a positive result i will ask to talk to their boss. \\
            3. & Eris sat in the interrogation room wringing her hands. "I was just in the wrong place at the wrong time. There can't be any conclusive evidence against me." She though to herself. Detective Opal entered the room and shut the door behind her. She sat across the table from Eris and looked deep into her eyes. "Things aren't looking good for you." She started solemnly. \\ \midrule
        \multicolumn{2}{p{15.5cm}}{\textbf{Dead Doctor} - \textit{Doctor is dead. She is the murder that the detective is trying to solve.}} \\ 
            1. & Lying on the floor, not much I can do. They think I'm dead, but I'm not. They should know this, they checked my pulse, I can't be dead. Am I dead? As I wonder, I see my whole life replay back before my eyes, it felt like years, but in reality it was only 20 minutes, and I regained consciousness, but couldn't move. \\ 
            2. & Having fought and clawed at her attacker until the bitter end she had played her role in helping to solve her own murder. The evidence was their hiding in plain sight beneath her darkly painted fingernails. Microscopic clumps of fur were embedded into her cuticles just waiting to plucked and analyzed. \\
            3. & I lie quietly in the morgue, not moving or seeing or breathing. The slab is quite cold and it is terribly quiet. The detective and the sergeant had stopped by earlier, but even though the Sergeant was a werewolf, he hadn't noticed and just thought I was another dead doctor. I am a Class F supernatural, and there are few who can recognize me -- for I am the Queen of the Vampires. I wait for night to fall when I will make my next move and begin hunting them down one by one. \\
    \bottomrule
    \end{tabular}
    \caption{An ideation example created by P2. Three characters are involved and their descriptions are listed accordingly.}
    \label{tab:example}
    \vspace{-0.2cm}
\end{table*}

\subsection{How Did the Participants Use \system?}

To capture how P1 and P2 used the system, we plotted the evolution of the cumulative word count to visualize their writing progress, aligned with the time they requested ideas.
Figure~\ref{fig:activity_log_1} and ~\ref{fig:activity_log_2} show the logs from P1 and P2, respectively. 
Both participants usually requested ideas and paused writing, which might signal getting stuck.
After a few hours, the participants came back, read all the returned story ideas and continued writing.
We also asked participants about how they interacted with \system in the post-study interview. Both participants wrote sequentially without any outlines. 
\begin{quote}
    \say{... I would write until I didn't know what to do next and then I would use the tool. The next day, I would read over everyone's responses and then write until I got stuck and then use the tool.} (P1)
\end{quote}
P2 finished all the writing within a day, so she tried various lengths of story prompts and launched several requests at the same time, as shown in Figure~\ref{fig:activity_log_2}.
Note that we allowed the participants to write freely (see the ``Study Protocol'' Section) and did not enforce any writing processes.

\subsection{Findings}
We summarize our findings of the study below, supported by quotes from the participants.

\textbf{The output of \system is interesting.}
Both P1 and P2 expressed that the ideas are interesting and fun:

\begin{quote}
    \say{Yeah, there are some very, very creative answers in there. Some people would just be like ... ``well, if I was this character, I would do this, this, and this.'' ... Some people would write a whole paragraph continuing the story. And I thought that was really interesting.} (P1)
\end{quote}

\begin{quote}
    \say{I really like it; it's pretty fun... that it came up with interesting stuff. There's one... ``Oh my gosh, that weirdo. I don't like her, booo.'' And it's just so funny... One of my favorite comments was like, [P2 read one idea] I was like, Oh, that's really interesting.... I thought that was really fun.} (P2)
\end{quote}

\textbf{\system is useful in generating inspiration.} Both P1 and P2 think \system can be useful for getting inspiring ideas from the crowd.

\begin{quote}
    \say{Yeah, it's helpful, even if I don't use their ideas.} (P1)
\end{quote}

\begin{quote}
    \say{It was nice when I got stuck on what to do next to be able to ask people. That gave me more inspiration to continue and also more insight into what people's expectations were for the story.} (P1)
\end{quote}

P2 said that, as an idea generator, \system produces relevant story ideas:

\begin{quote}
    \say{It's like an idea generator. I was really surprised by how much the ideas were actually related to the story.} (P2)
\end{quote}

P2 also mentioned that she would like to use the system for her next NaNoWriMo:

\begin{quote}
    \say{NaNoWriMo is coming up ... where you have to write a whole novel in a month ... sometimes it can be tricky to come up with ideas.} (P2)
\end{quote}

\textbf{Writers benefit from \system in different ways.}
We noticed that, although both participants think the system is useful, the way they used it was slightly different.
When P1 did not have ideas, people helped her figure out how to proceed, even inspiring the next part of the story.

\begin{quote}
    \say{Very helpful for when I don't have any ideas ... then I can ask a lot of other people and they'll help me figure it out. Even if I don't take any of their ideas, ... it might inspire something else that I will think of for the next part of the story.} (P1)
\end{quote}

P1 also agreed that \system can capture the personality of the role that is assigned to it.

\begin{quote}
    \say{Again, I think it depends on how much information the writer ... gives the people taking on the roles. And I think it can also (help) even if people don't get the correct personality, it still helps you learn ... ``well, they wouldn't do that, but they would do this instead.'' So it's still it's helpful to rule things out in that way too.} (P1)
\end{quote}

For P2, the ideas can be relevant to the characters: either they matched the character's personality or figured out a personality that had not been provided by P2.

\begin{quote}
    \say{...There's also... some things that are actually relevant to the characters. One of the characters ... was very dramatic. And then it (\system) came up with this idea that she would go make a lot of money and go to Vegas and like, sip martinis on an island somewhere... That's exactly like that!} (P2)
\end{quote}

\begin{quote}
    \say{There was another one... the surgeon says ``I'm going to do nothing. Except taking aspirin for my headache.'' And I was like, wait, \system remembered that he had a headache!... I didn't give them that much to go on. And it still had some personality figured out.} (P2)
\end{quote}

Not all the ideas was used in the stories, but they were still considered useful by participants because they inspired new thought.
For example, P2 specified in the role profile that the character ``Dead Doctor'' is a human, but a worker wrote an idea saying this character is a ``supernatural'' (the last row in Table 4).

\begin{quote}
    \say{(the worker) decided that one of my characters was a supernatural when I had said they were human. But this (the idea of setting ``Dead Doctor'' as a ``supernatural'') sounds like that's an interesting way of doing this.} (P2)
\end{quote}

Although P2 did not adopt this idea, this idea gave her story a new interesting direction to go. Such ideas stimulated participants ideas and thus were still considered useful.

\begin{figure}[t]
    \centering
    \includegraphics[width=0.95\columnwidth]{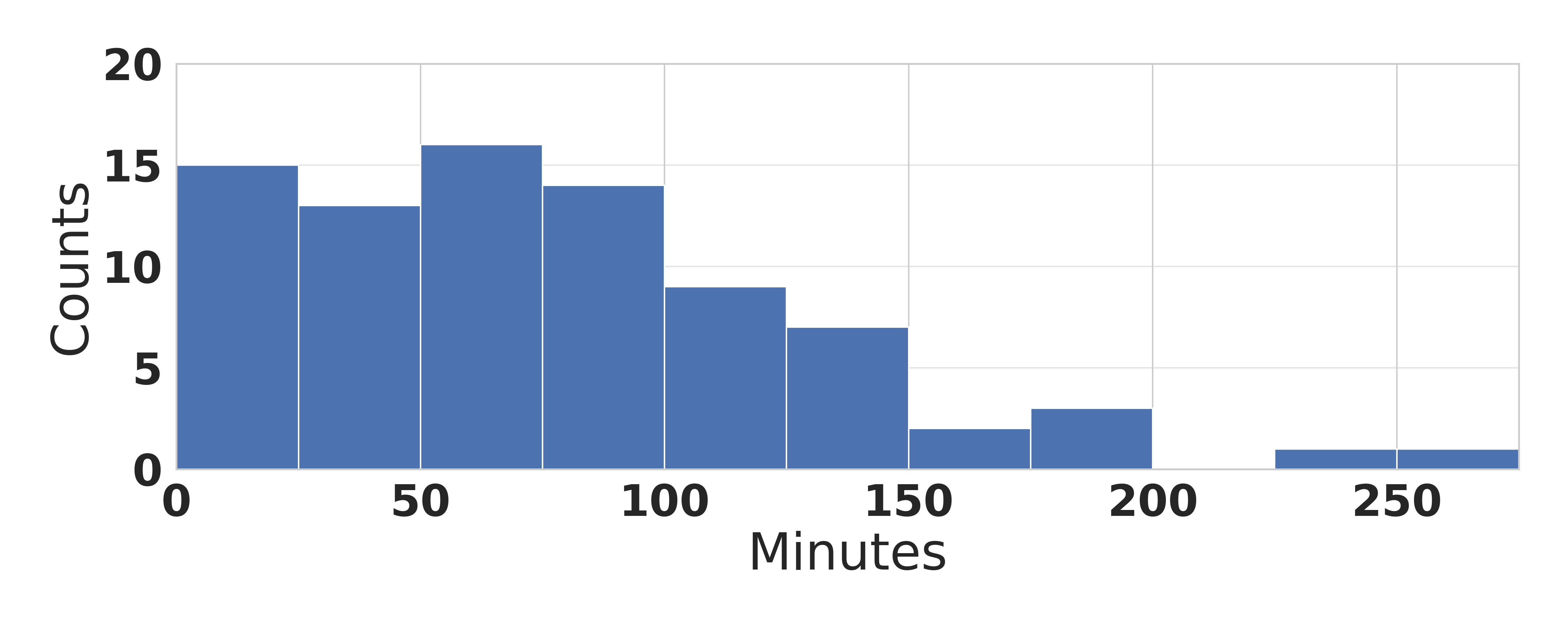}
    \caption{The histogram of the latency for getting responses from \system. A total of 81 responses were collected.}
    \label{fig:latency}
    \vspace{-1pc}
\end{figure}

\textbf{System latency did not greatly affect writing.} 
We investigated the system latency statistic and the resulting impact on users.
System latency is defined as the duration between the time the overview comment was automatically created and the time the story idea was received.
A histogram of system latency is shown in Figure~\ref{fig:latency}.
The latency for getting the first idea was around 15 minutes (median=14.42; mean=23.64; SD=25.14).
For each character to get at least one idea, the latency was about 50 minutes (median=53.32; mean=56.67; SD=30.47).
The latency for getting the last idea was about 160 minutes (median=167.12; mean=167.68; SD=52.72). 

We asked the participants if the system latency disrupted their writing processes. Both P1 and P2 said that the fact that \system needs a certain amount of time to respond did not affect their writing.
\begin{quote}
    \say{... because I had other things that I was doing. So, I would write, and then I would … do the other things during the day that I needed to do... I kind of had a schedule so it didn't really affect my time.} (P1)
\end{quote}

\begin{quote}
    \say{Probably, but it wasn't too bad... when you're coming up with stuff, it's always ... a long process anyway... it's not really like, ``okay, I got the whole idea down now'' ... some of the ideas, it's kind of like ``that's really interesting; maybe I'll go back and change it.'' But it's not that big of a deal to go back and change it because you have to ... make like four or five drafts anyway.} (P2) 
\end{quote}

\textbf{The role play strategy does not fit some use cases.} Both participants pointed out some problems they encountered, some of which were caused by the nature of the role play strategy. P1 would like to use the baseline strategy (no-role) and dynamic team management in some cases:

\begin{quote}
    \say{I'm conflicted between wanting to be able to assign a certain task to just one character versus the whole team because sometimes characters don't fall in... I would imagine if you have a larger cast of characters, the teams would overlap quite a bit, so it might be easier to be able to... assign tasks to single characters.} (P1)
\end{quote}

P1 also pointed out that some scenarios might be hard to use the role play strategy, since the structure of the story will be too complex:

\begin{quote}
    \say{... it depends on how detailed you are when you write the character, because I only wrote a couple sentences for my characters, but if you wrote ... a whole biography, then maybe. I think it depends on the complexity of the story and the complexity of the characters.} (P1)
\end{quote}

\textbf{Working with stranger workers have trade-offs.}
We asked the participants to compare working with strangers versus working with friends, families, or colleagues. P1 explained the trade-offs between them.

\begin{quote}
    \say{(Using \system) It's less pressure because you don't know the people, but it's also a little more nerve-wracking because you don't know the people. So there's good and bad... It's better to have someone that you know and have a good relationship with. It's hard to trust strangers with a story, especially a story as complex as a book.} (P1)
\end{quote}

Copyright issues were also raised if users were to use \system for their own professional work.

\begin{quote}
    \say{I also think that if a professional writer was going to use the tool in their professional work, it might raise copyright issues. ... If you are getting ideas from other people, and you implement them in a book that you're going to sell, who gets the credit for them?} (P1)
\end{quote}

\textbf{Handling overwhelming number of story ideas.} Both participants thought the number of ideas provided by \system was overwhelming.

\begin{quote}
    \say{Depending on how big it gets, it might be overwhelming to have to read through all those responses.} (P1)
\end{quote}
    
\begin{quote}
    \say{It's not that there are too many it's just that I didn't realize how many.} (P2)
\end{quote}

%% file: discussion.tex
In this section we discuss topics that are broader than the scope of the \system system.

\subsection{Differences Between Technical and Creative Writing}
P2 in Experiment 2 had experience in both technical writing and creative writing. 
In order to better inform our future system design, we asked P2 in pre-study interview what are differences between these two.
P2 said she thinks technical writing is much easier because the goal and style is more clear.


\begin{quote}
    \say{Often (in technical writing) you have a style guide, and you have the a goal. The goal with technical writing is to make something that's confusing understandable. Whereas the goal with creative writing is usually to give some kind of feeling to the reader.} (P2)
\end{quote}

She further explained that, in creative writing, the writers sometimes need to intentionally avoid clear explanations, whereas technical writing is all about making things clear and understandable.

\begin{quote}
    \say{So, with technical writing, you're just explaining something, you're trying to make something very clear. But sometimes with creative writing, you might not necessarily want to be making something clear. You might want to introduce moral quandaries to your reader and make them think about all of the gray areas and like ... how it's not as clear as you thought, I guess. \textbf{That's a big, big difference.}} (P2)
\end{quote}

\vspace{-0.1cm}
\subsection{What Do Writers Do to Resolve Writer's Block}
P1 struggles with plot the most. She said that knowing the characters more can help with the situation since characters and plots are intertwined.

\begin{quote}
    \say{Plot is my weakest skill when it comes to writing... So I find that if I really get to know my characters really well and understand the choices that they would make in any given situation, then the plot can kind of unravel itself from there.} (P1)
\end{quote}

P2 usually let the characters talk when getting stuck.
Even if the conversations are deleted afterwards, it helps her understand the character more.

\begin{quote}
    \say{When I get stuck, what I usually do is just have characters talk about dumb stuff... just have two characters talk to each other... And then sometimes you figure out more about what you want to do by that conversation... It helps you understand the characters more, if you have them talk to each other. And then knowing ``now I know that this person wants to do this.''} (P2)
\end{quote}

\vspace{-0.2cm}
\subsection{How Do Writers Write}
P1 stated that her writing process was to come up with an idea, create characters, and finally, design a plot.
P1 also said that she would write one draft first and later revise for plot and character. 

\begin{quote}
    \say{The idea always comes first. So I always have ``what if this happened; that would be an interesting story.'' And then I create the characters for that idea. The plot comes last. I will write one draft and then revise for plot and character. But it's always idea, characters, plot.} (P1)
\end{quote}

P2 usually thinks about the message she wants to send, picks overall ``concepts of things,'' thinks about characters and sets up the inner/outer goals, and figures out what she needs in the plot to satisfy the characters' needs.

\begin{quote}
    \say{So first, I think about what message I want to send. What ... is something that I want to talk about or discuss? And I'll pick ... the setting and the time period. You know, overall concepts of things like social change that I want to talk about. Then I'll think about the character. Because for me, the character drives the plot. ... Usually characters have two goals. The first is the outer goal and the second is an inner goal. ... You figure out what you want in the plot based on how it'll satisfy the characters' needs.} (P2)
\end{quote}

\vspace{-0.2cm}
\subsection{Limitations}

\textbf{The need of non-role strategy.} 
\system currently only supports a role-playing strategy, so we do not observe any cases where the user prefers to request ideas without role schema.
We will add new features to \system to allow users to request ideas using different strategies.

\textbf{Scalability.}
Our experiment focused on short stories (under 1,000 words) with a few characters who have relatively simple backstories.
When working on long stories, the structures and characters may become complicated, raising two issues: handling a large number of characters and conveying complicated backstories.
\system currently requires users to define characters and teams before writing stories. However, when a story has more characters, it can become difficult to handle all the characters and teams.
Features such as automatically suggesting teams based on context might help. We will also explore automatic summarization technologies to produce or update character backstories automatically.

\textbf{Insufficient amount of participants.}
Only two participants were recruited in this study, as it is hard to recruit creative writers who are willing to participate in a multiple-day study.
A one-day study might be too short for creative writers to come up with good ideas for stories. In the future, a deployment several months long with more users would allow us to better understand how people interact with \system.

\textbf{Evidence for relieving writer's block.}
In this paper, we showed that the role-playing strategy produces semantically far story plot ideas (Study 1), and participants were satisfied with the ideas provided by \system (Study 2).
However, we did not directly examine whether the system helped relieve writer's block. Evaluating the usefulness of an idea is challenging because an idea can still be considered useful or inspiring even if it is not directly adopted.
A large-scale deployment will allow us to observe whether writers use the ideation feature frequently, which could better validate the usefulness of \system.

%% file: conclusion-and-future-work.tex
This paper introduces \system, a crowd-powered system that applied crowd ideation to help creative writers.
We built \system as a Google Docs add-on, and writers can simply use the editor to elicit story ideas with the online crowd.
\system adopts the role play strategy for story ideation.
In controlled experiments, this strategy produced story ideas that are semantically more distant to the working story draft, which is known to be more useful to creator who reaches an impasse.
We also conducted a deployment study with two experienced creative writers.
In the deployment study, we found that the outputs of \system is generally interesting and useful, while two participants benefit from the system in different ways.
In the future, we will relax the definition of ``characters'' in our system, allowing writers to use these ``roles'' in \system in a more general way. For example, each role could be a ``thinking hat'' that represents one perspective~\cite{de2017six}.